\documentclass[11pt, a4paper]{article}
\pdfoutput=1

\usepackage{amsmath}
\usepackage{amssymb}
\usepackage{comment}
\usepackage{multirow}
\usepackage[utf8]{inputenc}
\usepackage{bm}
\usepackage{cite}
\usepackage{physics}
\usepackage{slashed}
\usepackage{braket}
\usepackage{enumitem}
\usepackage{xargs}
\usepackage{tcolorbox}
\usepackage{xfrac}
\usepackage[footnotesize]{caption}
\usepackage{fourier}
\usepackage{cancel}
\usepackage{soul}
\usepackage{hyperref}

\definecolor{rossoferrari}{HTML}{D9073D}
\definecolor{mediumblue}{HTML}{0000CD}
\definecolor{forestgreen}{HTML}{228B22}
\definecolor{desy_blue}{HTML}{009EE2}
\definecolor{desy_orange}{HTML}{FD8800}
\definecolor{light_pink}{rgb}{1,0.4,0.4}
\definecolor{light_blue}{rgb}{0.284602,0.317763,0.963947}
\hypersetup{% hyperref option list
setpagesize=false,
bookmarksnumbered=true,%
bookmarksopen=true,%
colorlinks=true,%
linkcolor=light_blue,
urlcolor=forestgreen,
citecolor=forestgreen,
linktocpage=false,
}

\usepackage[height=8.85in,width=6.8in]{geometry}

\makeatletter
\@addtoreset{equation}{section}

\makeatother

\renewcommand{\vec}[1]{\bm{#1}}

\begin{document}
\title{Electromagnetic (high-frequency) gravitational wave detectors: \\ Interferometers revisited}

\author{ Valerie Domcke$^a$ and Joachim Kopp$^{a,b}$ }
\date{}

\maketitle

\begin{center}
{\small
\begin{tabular}{ll}
$^a$&\!\!\!\!\!\! \emph{Theoretical Physics Department, CERN, 1211 Geneva 23, Switzerland}\\[.3em]
$^b$&\!\!\!\!\!\! \emph{PRISMA+ Cluster of Excellence \& Mainz Institute for Theoretical Physics, 55128 Mainz, Germany}
\end{tabular}
}
\end{center}

\begin{abstract}
Increased interest in pushing the frontier of gravitational wave searches to higher frequencies (kHz and beyond) has resulted in a variety of different proposed experimental concepts. A significant fraction of them are based on the coupling between classical electromagnetism and gravity. We highlight some differences and similarities between different approaches, showcasing the rich phenomenology arising from this coupling. We use the opportunity to re-derive the response function of interferometers as the low-frequency limit of a broader picture. This article was prepared as proceedings for  Moriond Cosmology 2024.
\end{abstract}

% 8 pages

%-----------------------------------------------------------------------------
\section{Introduction}
%-----------------------------------------------------------------------------

Since the first direct observation of gravitational waves (GWs) by the LIGO--Virgo collaboration in 2015~\cite{LIGOScientific:2016aoc}, gravitational wave astronomy has grown into a field of many opportunities, providing a new avenue to explore our Universe. Established searches span a wide range of frequencies, from the very low frequencies probed by the CMB, over the nHz range targeted by pulsar timing arrays and the mHz range which is the goal of the upcoming LISA mission, up to 100~Hz as explored by ground-based interferometers. Even higher frequencies are less well explored, but could eventually
%mostly due to a lack of known astrophysical sources above kHz.  Nevertheless, GWs from MHz to THz and beyond could 
probe a multitude of scenarios beyond the Standard Model, including for instance phase transitions in the early Universe, primordial black holes, and cosmic strings~\cite{Aggarwal:2020olq}. While the sensitivity of current high-frequency GW detectors still falls several orders of magnitude short of testing realistic scenarios, rapid progress is being made. Interestingly at high frequency, the landscape of proposed detectors is very diverse~\cite{Aggarwal:2020olq}. Among the leading competitors are concepts exploiting the GW coupling to electromagnetism. Moreover, the mechanical coupling of GWs, at the basis of the original Weber bar experiments~\cite{Weber:1967jye}, has seen a revival~\cite{Goryachev:2014yra,Aggarwal:2020umq,Berlin:2023grv}. But also more exotic processes, such a the GW spin coupling, are being explored~\cite{Ito:2022rxn}.
%Across this entire frequency range, it has been, loosely speaking, the coupling of GWs to electromagnetism which has proven to be the most promising avenue for GW detection. As we push the frontier to even higher frequencies, the landscape of existing and proposed detectors becomes more diverse.

Here, we focus on electromagnetic detection of (high-frequency) GWs. In this context, it has proven useful to describe the impact of the GWs as an `effective current' or `effective polarization and magnetization' which appears in Maxwell's equations when evaluated in a GW background. In these proceedings, we demonstrate explicitly how to obtain the textbook results for interferometer response functions in this language. Instead of evaluating the change in proper distance between the two end-points of an interferometer arm, we explicitly solve  Maxwell's equations in presence of the GW-induced effective source term. Not surprisingly, we recover the familiar result for the phase shift (i.e.\ the change in proper distance) in the limit that the GW frequency is small compared to the photon frequency, as obtained previously for the special case of orthogonal GW and photon directions. We discuss additional properties of the full solution, and their relevance for (high-frequency) GW searches~\cite{Melissinos:2010zz}.

As a disclaimer, we stress that neither the methods presented here nor the results are new (in fact, both are decades old).  We find it nevertheless insightful to explicitly demonstrate a range of effects resulting from a single, simple computation, as opposed to the range of methods that have been previously employed to understand different aspects.

%-----------------------------------------------------------------------------
\section{Maxwell's equations in curved space time}
%-----------------------------------------------------------------------------

Consider a Minkowski background $\eta_{\mu\nu} = {\rm diag}(-, +, +, +)$,  perturbed by a GW, $g_{\mu \nu} = \eta_{\mu \nu} + h_{\mu \nu}$.  In the absence of external currents, the inhomogeneous Maxwell equation for the electromagnetic field strength tensor $F_{\alpha \beta}$ reads~\cite{Landau:1975pou}
\begin{equation}
 \nabla_\mu ( g^{\mu \alpha} F_{\alpha \beta} g^{\beta \nu})  = \partial_\mu (\sqrt{-g} \; g^{\mu \alpha} F_{\alpha \beta} g^{\beta \nu}) = 0 \,.
\end{equation}
Expanding to linear order in the GW amplitude $h$ yields
\begin{equation}
 \partial_\mu F^{\mu \nu} + \partial_\mu (\frac{1}{2} h F^{\mu \nu} - h^{\mu \alpha} F_{\alpha}^{\; \nu} -  F^{\mu}_{\; \beta} h^{\beta \nu}) = 0
\end{equation}
where $F_{\mu \nu} = \partial_\mu A_\nu - \partial_\nu A_\mu$, $h = h^\mu_{\ \mu}$, and we raise and lower indices with $\eta_{\mu \nu}$. We can now expand $F_{\mu\nu} = \bar F_{\mu\nu} + {\cal F}_{\mu\nu}$ with $\bar F_{\mu\nu}$ solving Maxwell's equations in flat space, $\partial_\mu \bar F^{\mu\nu} = 0$, and ${\cal F}_{\mu \nu}$ denoting the EM field induced by the GW. With ${\cal F}_{\mu\nu} = {\cal O}(h)$, this yields to linear order in $h$,
\begin{equation}
0 = \partial_\mu {\cal F}^{\mu \nu} +  \partial_\mu ( \tfrac{1}{2} h \bar F^{\mu \nu} - h^{\mu \alpha}  {\bar F}_{\alpha}^{\; \nu} -  \bar F^{\mu}_{\; \beta} h^{\beta \nu}) \equiv  \partial_\mu {\cal F}^{\mu \nu} - j^\nu_{\rm eff}  \,,
 \label{eq:MW}
\end{equation}
i.e.\ the GW, together with the background EM field $\bar F$, acts as an effective current $j^\nu_{\rm eff}$ (or equivalently an effective polarization and magnetization~\cite{Domcke:2022rgu}) sourcing the induced EM field ${\cal F}_{\mu \nu}$. 
The homogeneous Maxwell equation is not affected by the GW.  

In the following, we will focus on the response of GW detectors to a gravitational plane wave in the transverse traceless frame,
\begin{align}
h_{ij}(\vec x) = (h_+ e_{ij}^+(\hat{\vec k}_g) + h_\times e_{ij}^\times (\hat{\vec k}_g))  \cos[ \omega_g (t - \hat{\vec k}_g \cdot  \vec x)] \,,  \quad h_{\mu 0} = h_{0 \mu} = 0\,,
\end{align}
with $\hat{\vec k}_g = (\sin\theta \cos\phi, \sin\theta \sin\phi, \cos\theta)$ and $\omega_g = |\hat{\vec k}_g|$ denoting the wave vector and frequency of the GWi, respectively. The polarization tensors 
\begin{align}
\hat e_{ij}^+ = u_i u_j - v_i v_j \,,  \quad \hat e_{ij}^\times = u_i v_j + v_i u_j  \quad \text{with} \quad    \vec v = \hat{\vec e}_\phi = (-s_\phi, c_\phi, 0) \,,  \quad \vec u = \vec v \times \hat{\vec k}_g \,,
\end{align}
are normalized to $e_{ij}^\lambda e^{ij}_{\lambda'} = 2 \delta_{\lambda \lambda'}$. We will use $s_\phi$ and $c_\phi$ to abbreviate $\sin \phi$ and $\cos \phi$, respectively. 
The transverse traceless frame is particularly well suited to describe free-falling objects (such as the mirrors in an interferometer) or to describe the high-frequency limit of any material response, well above the relevant mechanical resonance frequencies. For a detailed discussion on different frame choices and their equivalence, see~\cite{Ratzinger:2024spd}.

%-----------------------------------------------------------------------------
\section{Towards interferometers: a single link}
\label{sec:1arm}
%-----------------------------------------------------------------------------

Let us start by considering a single link in an interferometer, i.e.\ a linearly polarized EM wave with vector potential $\vec{A}$ (in temporal gauge, $A^0 = 0$) and angular frequency $\omega_\gamma$ propagating along the $x$-axis, 
\begin{equation}
 \vec A = \bar{ \vec A} + \vec{{\cal A}} = A_0 \cos[\omega_\gamma (t - x)]  \hat e_z  + \vec{{\cal A}} \,,
 \label{eq:A}
\end{equation}
where as before $\bar{\vec{A}}$ denotes the solution to Maxwell's equations in flat space and $\vec{{\cal{A}}}$ is the GW-induced contribution. Given that we can express the source term in Eq.~\eqref{eq:MW} as a linear combination of 
\begin{align}
 \cos_+ \equiv \cos[ (\omega_\gamma + \omega_g) t - \omega_\gamma x - \hat{\vec k}_g \cdot \vec x  ]
 \quad \text{and} \quad
 \cos_- \equiv \cos[ (\omega_\gamma - \omega_g) t - \omega_\gamma x + \hat{\vec k}_g \cdot \vec x ] 
\end{align}
we proceed by choosing the ansatz
\begin{align}
 {\cal A}_i = a_i^- \cos_- + a_i^+ \cos^+ . 
\end{align}
Inserting this into Eq.~\eqref{eq:MW} and taking the GW to be in the $x-z$ plane at an angle $\vartheta = \pi/2 - \theta$ from the $x$-axis, we obtain 
\begin{align}
 a_x^\pm & = A_0 h_+ s_\vartheta \frac{(\omega_g \pm \omega_\gamma(1 - c_\vartheta))}{4 (\omega_g \pm \omega_\gamma)} 
  = \frac{1}{4} A_0 h^+ s_\vartheta (1 - c_\vartheta) + {\cal O}(\omega_g/\omega_\gamma) \,, \\
 a_y^\pm & =  \tfrac{1}{4} A_0 h_\times  \,, \\[0.15cm]
 a_z^\pm & = - A_0 h_+ \frac{\omega_\gamma}{\omega_g} \left( 8 \omega_\gamma (\omega_g \pm \omega_\gamma) \right)^{-1} 
  \cdot \left( 2  c_\vartheta(\omega_g^2 \pm \omega_g \omega_\gamma + \omega_\gamma^2) \pm \omega_\gamma(\omega_g \pm 2 \omega_\gamma - \omega_g c_{2 \vartheta}) \right)\nonumber \\ 
 &= \mp \frac{1}{4} A_0 h^+ \frac{\omega_\gamma}{\omega_g} (1 + c_\vartheta) + {\cal O}[(\omega_g/\omega_\gamma)^0] \,. 
 \label{eq:az0}
\end{align}
For applications at interferometers, $\omega_g/\omega_\gamma$ is tiny and hence we can drop all but the leading order terms in this ratio. Noting that only $a_z^\pm$ (or, more generally, the component of $\vec{{\cal A}}$ which corresponds to the polarization of the background EM wave) features a term \textit{enhanced} by a factor of $\omega_\gamma / \omega_g$, we will mainly focus on this component in the following.

The solution above is a particular solution of Eq.~\eqref{eq:MW}. The general solution is obtained by adding plane waves, and the physical solution of interest is the one that obeys the appropriate boundary conditions. For the application at interferometers, we impose the boundary condition $\vec{{\cal A}}(t, x = 0) = 0$. At leading order in $\omega_g/\omega_\gamma$ this can be achieved by adding plane waves $\alpha_z^\pm \cos[(\omega_\gamma \pm \omega_g)(t - x)]$ with amplitude
\begin{align}
 \alpha_z^\pm  & = \pm \frac{A_0 h^+ \omega_\gamma}{4 \omega_g} (1 + c_\vartheta) \,.
\end{align}
Combining the above results we thus obtain the solution
\begin{align}
\vec{{\cal A}}(t, \vec x) = & \frac{A_0 h^+ \omega_\gamma}{4 \omega_g} (1 + c_\vartheta) \, \hat{\vec e}_z  \, \times \nonumber \\
&  \hspace{-0.8cm} \sum_{\lambda = \pm} \lambda \left(  \cos[(\omega_\gamma + \lambda \omega_g)(t - x)]  - \cos[ (\omega_\gamma + \lambda \omega_g) t - \omega_\gamma x -  \lambda  \hat{\vec k}_g \cdot \vec x ] \right)   + {\cal O}\big([\omega^g/\omega^\gamma]^0\big),
\label{eq:Asol}
\end{align}
for the EM wave induced by a GW crossing a background EM wave at an angle $\vartheta$. This result has been obtained previously for the special case of $\vartheta = \pi/2$~\cite{Melissinos:2010zz}, as well as for general incident angles~\cite{Bringmann:2023gba}.

In the context of interferometers, we are interested in the phase shift accumulated by the propagating EM wave due to the presence of the GW. Given Eq.~\eqref{eq:Asol} this can now be read off as
\begin{align}
  \Delta \phi(t)  = \arctan[ - A_z(t)'/ A_z(t)] - \arctan[ - \bar A_z(t)'/ \bar A_z(t)]    = \frac{\bar A_z' {\cal A}_z - \bar A_z {\cal A}_z'}{\bar A_z^2 + \bar A_z'^2} + {\cal O}({\cal A}_z^2/A_z^2)\,,
\end{align}
where the prime denotes the derivative with respect to $\omega_\gamma t$. This yields 
the well-known result
\begin{align}
 \Delta \phi = - \frac{1}{2} h_+ \omega_\gamma L  \sin^2 \vartheta + {\cal O}(\omega_g/\omega_\gamma) \,.
  \label{eq:2DLigoArm}
\end{align}
for the phase shift of a photon travelling a distance $L$ in the limit $\omega_g L \ll 1$. This result is more commonly (and admittedly more rapidly) obtained by evaluating the change in proper distance between two test-masses induced by the GW, leading to a corresponding time-delay (i.e.\ phase shift) for an EM wave propagating between these two test masses~\cite{Maggiore:2007ulw}, reproduced for convenience in the appendix. As expected, we reproduce this result at leading order in $\omega_g/\omega_\gamma$, but note that the full solution for the induced EM wave obtained above allows us to read off further properties of the induced signal, such as the amplitude variation and a rotation of the photon polarization (both at order $(\omega_g/\omega_\gamma)^0$).

A few comments are in order. In the discussion above, we have dropped phase factors for the GW and EM wave at $(t,\vec x) = (0, \vec 0)$. They can easily be restored by replacing  $\omega_\gamma t \mapsto \omega_\gamma t + \phi_\gamma$ and $\omega_g t \mapsto \omega_g t + \phi_g$ in the arguments of the cosines, respectively. We have also taken the photon to be polarized in the plane spanned by the GW and photon wave vector. Repeating the exercise for a photon polarized in the $y$-direction, we obtain
 \begin{align}
 a_x^\pm &   = \frac{1}{2} A_0 h^\times  s_\vartheta + {\cal O}(\omega_g/\omega_\gamma) \,, \\
 a_y^\pm & =   \mp \frac{1}{4} A_0 h^+ \frac{\omega_\gamma}{\omega_g} (1 + c_\vartheta) + {\cal O}[(\omega_g/\omega_\gamma)^0]   \,, \\
 a_z^\pm & = - \frac{1}{4} A_0 h^\times (1 + 2 \cos \vartheta)  + {\cal O}(\omega_g/\omega_\gamma) \,.
\end{align}
This demonstrates that the leading order contribution (here in $a_y^\pm$) appears in the component of $\vec{{\cal A}}$ parallel to the background EM wave polarization and does not depend on the orientation of the gravitational wave vector with respect to this direction. Finally, the absence of the $\times$-polarization in the leading order result is a geometric projection effect due to the EM wave propagating in the $x$-direction in combination with a GW vector in the $x-z$ plane. As we will see below, we will recover the phase shift induced by the $\times$-polarization in a more general setup.

%-----------------------------------------------------------------------------
\section{Interferometer response functions}
%-----------------------------------------------------------------------------

In this section we will generalize the results obtained above to a general GW incidence angle, enabling us to reproduce the response functions of GW interferometers such as LIGO. Given the above observations, we will consider without loss of generality (at leading order in $\omega_g/\omega_\gamma$)  a background EM wave polarized in $z$ direction, and focus solely on the equation of motion for ${\cal A}_z$,
\begin{align}
 \partial_x^2 {\cal A}_z - \partial_t^2 {\cal A}_z + {\cal S}_z^+ +  {\cal S}_z^\times = 0 \,,
\end{align}
where the source term ${\cal S}_z$ depends on the orientation of the interferometer arm which we take to be the $x$- or $y$-axis,
\begin{align}
 {\cal S}_z^+ =  A_0 h_+ \omega_\gamma^2 \begin{cases}
  (c_\theta^2 c_\phi^2 - s_\phi^2)  \cos[(\omega_\gamma (t \pm x)]\cos[\omega_g(t - \hat{ \vec k}_g \cdot \vec x)]  & \text{for $x$-arm} \\
  (c_\theta^2 s_\phi^2 - c_\phi^2) \cos[(\omega_\gamma (t \pm y)]\cos[\omega_g(t -   \hat{ \vec k}_g \cdot \vec x)]  & \text{for $y$-arm} 
 \end{cases}\,,
 \label{eq:+source}
\end{align}
and 
\begin{align}
 {\cal S}_z^\times  = A_0 h_\times \omega_\gamma^2 \begin{cases}
 - 2 c_\theta c_\phi s_\phi  \cos[(\omega_\gamma (t \pm x)]\cos[\omega_g(t -  \hat{ \vec k}_g \cdot \vec x)]  & \text{for $x$-arm} \\
  c_\theta s_{2 \phi}  \cos[(\omega_\gamma (t \pm y)]\cos[\omega_g(t -   \hat{ \vec k}_g \cdot \vec x)]  & \text{for $y$-arm}
 \end{cases} \,.
\end{align}
Here $\theta$ and $\phi$ denote the orientation of the GW vector in spherical coordinates, note in particular that (contrary to $\vartheta$ above), $\theta$ is defined as the angle relative to the $z$-axis.  In the limit $\omega_g L \ll 1$, relevant to ground-based interferometric detectors, the second trigonometric factor only contributes at higher order in $\omega_g L$ so that we can read off the generalization of Eq.~\eqref{eq:2DLigoArm} by comparing the angular dependence of the prefactors in ${\cal S}^\pm_z$ with the analogous expression
\begin{align}
{\cal S}_z = A_0  h_+ \omega_\gamma^2 s_\vartheta^2 \cos[\omega_\gamma(t - x)] \cos[\omega_g (t -  \hat{ \vec k}_g \cdot \vec x)]
\end{align}
obtained for the setup of the previous section (which corresponds to the limit $\phi = 0$, $\theta = \pi/2 - \vartheta$ of the first line in Eq.~\eqref{eq:+source}). Note that all angular prefactors are identical under inversion of the photon direction, implying that both legs of the return trip contribute equally to the phase shift. This yields the required geometric projection factors for a GW with general incident angle, and hence we directly obtain the appropriate phase shifts by replacing $\sin^2\vartheta$ in Eq.~\eqref{eq:2DLigoArm} accordingly. 

For example, for the $+$-polarization, we replace $\sin^2\vartheta \mapsto c_\theta^2 c_\phi^2 - s_\phi^2$ for a photon propagating in $x$-direction. This gives the total phase shift accumulated in this arm,
\begin{align} 
\Delta \phi_x^+ = - (c_\theta^2 c_\phi^2 - s_\phi^2) h_+ \omega_\gamma L \,.
\end{align}
Similarly, for a photon propagating in $y$-direction,  we replace $\sin^2\vartheta \mapsto c_\theta^2 s_\phi^2 - c_\phi^2$, yielding
\begin{align} 
\Delta \phi_y^+ = - (c_\theta^2 s_\phi^2 - c_\phi^2) h_+ \omega_\gamma L \,.
\end{align}
The phase difference is thus obtained as $\Delta \phi^+ = \Delta \phi^+_x -  \Delta \phi^+_y = - L \omega_\gamma h_+ (1 + c_\theta^2)c_{2 \phi}$. Repeating this exercise with the $\times$-polarization we obtain
\begin{align}
\Delta \phi_\text{LIGO} =  L \omega_\gamma \left( 2 h_\times c_\theta s_{2 \phi} -  h_+ (1 + c_\theta^2)c_{2 \phi} \right)\,.
\label{eq:LIGO}
\end{align}
This is the well-known result for the phase shift induced in LIGO by a passing GW, reproduced for convenience in the appendix using the standard derivation through the change in proper distance in the transverse traceless frame.

%-----------------------------------------------------------------------------
\section{Beyond interferometers - other electromagnetic GW detectors}
%\section{Effects at higher order in $\omega_g/\omega_\gamma$ }
%-----------------------------------------------------------------------------

%Comment in more detail on amplitude modulation and rotation of the polarization ? Gets potentially interesting at higher GW frequencies.

The previous section has focused on the impact of a GW on an EM wave in the limit $\omega_g \ll \omega_\gamma$, which is an excellent approximation for interferometers based on optical photons such as LIGO/Virgo/Kagra (operating in the 100~Hz range) and LISA (which will operate in the mHz range), as well as for pulsar timing arrays, utilizing radio photons for detecting nHz GWs. In these cases, the phase shift $\Delta \phi$, in particular when read out with an interferometric measurement, is the most accessible observable.

More generally, there are however additional effects, as the discussion above teaches us. For example, the leading order amplitude variation in the setup discussed in Sec.~\ref{sec:1arm} is 
\begin{align}
\Delta \langle A(t) \rangle^2 = 
 \frac{h_+}{4} A_0  \cos^2(\vartheta/2) \left( \cos[(t - x) \omega_g ] - \cos[(t - c_\vartheta x)\omega_g] \right) \,,
 \label{eq:Amod}
\end{align}
where $\langle \cdot \rangle$ denotes coarse-graining in time, that is, averaging over oscillations with frequencies of order $\omega_\gamma$. Similarly, the rotation of the photon polarization (obtained from the $(\omega_g/\omega_\gamma)^0$ terms in Eq.~\eqref{eq:az0}) is of order $h$, and will oscillate at the side-band frequency $\omega_\gamma \pm \omega$. Suggestions to search for the latter  effect have been put forward in \cite{Cruise:1983,Cruise:2000}.

When searching for GWs at higher frequencies the experimental concepts currently under consideration are more diverse,\cite{Aggarwal:2020olq} and the limit $\omega_g \ll \omega_\gamma$ does not always apply. It is thus crucial to have access to the full EM solution. For example, experiments based on the resonant conversion of GWs to photons (which is possible in axion helioscopes and light-shining-through-wall experiments~\cite{Ejlli:2019bqj,Ringwald:2020ist}) use a static background magnetic field, and momentum conservation dictates that the frequency of the generated photon matches the frequency of the incoming GW. The conversion probability of these experiments can be obtained by solving Eq.~\eqref{eq:MW} with $\vec j_\text{eff}$ determined by the static background magnetic field and the incoming GW. For astronomically large length scales (such as conversion in cosmic magnetic fields~\cite{Domcke:2020yzq} or neutron stars~\cite{Liu:2023mll,Ito:2023fcr,Dandoy:2024oqg}) the back-conversion from photons to GWs can become important, requiring to solve the coupled system of Eq.~\eqref{eq:MW} and the linearized Einstein equation for GW propagation including the EM fields as a source term~\cite{Raffelt:1987im}.
Another type of axion experiments that have been proposed as GW detectors are low-mass axion haloscopes which search for an axion- or GW-induced magnetic flux  using resonant LC circuits. The induced flux arises due to coupling of a static background magnetic field to a GW through the effective current in Eq.~\eqref{eq:MW}~\cite{Domcke:2022rgu,Domcke:2023bat}. As above, the frequency of the induced flux matches the GW frequency.  
On the contrary, in resonant cavity searches, the GW induces a transition between two modes of a superconducting radio-frequency (SRF) cavity~\cite{Berlin:2021txa,Berlin:2023grv}. Again the interaction is described by Eq.~\eqref{eq:MW}, and depending on the setup the GW frequency should be similar to or smaller than the frequency of the lowest cavity modes.  In all these experiments, the goal is to generate photons from GWs in the presence of a background EM field, and hence the observable is not a phase shift but rather the presence of these additional photons. 

Finally, atomic clocks have been proposed to measure the frequency modulation of photons propagating in a GW background~\cite{Bringmann:2023gba}. This application is safely in the $\omega_g \ll \omega_\gamma$ limit, and the frequency modulation is equivalent to the phase and amplitude modulation found in Eqs.~\eqref{eq:LIGO} and \eqref{eq:Amod}.

%-----------------------------------------------------------------------------
\section{Conclusions}
%-----------------------------------------------------------------------------

These proceeding highlight similarities and differences in GW searches based on the interaction of electromagnetism with gravity. While interferometers are by far the most advanced concept, alternative ideas are gaining momentum to push the frontier of GW searches to higher frequencies. While new theoretical tools are being developed and old tools are being rediscovered to describe this new experimental terrain, we demonstrate here explicitly how to recover textbook results for interferometers using the language frequently employed in the high-frequency GW community.

%-----------------------------------------------------------------------------
\section*{Acknowledgments}
%-----------------------------------------------------------------------------

It is a pleasure to thank our great collaborators on various projects exploring EM GW detectors, in particular Torsten Bringmann, Sebstian Ellis, Camilo Garcia Cely, Elina Fuchs, Sung Mook Lee and Nick Rodd.

%-----------------------------------------------------------------------------
\section*{Appendix}
%-----------------------------------------------------------------------------

For completeness and notational consistency, we recapitulate in this appendix the standard derivation of the response function of an interferometer~\cite{Maggiore:2007ulw}. In the transverse traceless frame, the proper distance between two test masses is given by
\begin{align}
ds^2 = - dt^2 + (\delta_{ij} + h_{ij})dx^i dx^j \,.
\end{align} 
From this, we obtain the phase shift for a photon propagating in $x$-direction as $\Delta \phi_x = -\tfrac{1}{2} h_{xx} L \omega_\gamma$ and analogously for the $y$-direction. The return trip adds a factor 2. This yields the total phase difference as
\begin{align}
\Delta \phi_\text{LIGO} = (h_{xx} - h_{yy}) L \omega_\gamma =  L \omega_\gamma \left( 2 h_\times c_\theta s_{2 \phi} -  h_+ (1 + c_\theta^2)c_{2 \phi} \right)\,,
\end{align}
which matches the result~\eqref{eq:LIGO} obtained in the main text. From this result, one can read off the interferometer response including the antenna pattern in the usual way.

\small
\bibliographystyle{utphys}
\bibliography{moriond24}

\providecommand{\href}[2]{#2}\begingroup\raggedright\begin{thebibliography}{10}

\bibitem{LIGOScientific:2016aoc}
{\bfseries LIGO Scientific, Virgo} Collaboration, B.~P. Abbott {\em et~al.},
  ``{Observation of Gravitational Waves from a Binary Black Hole Merger},''
  \href{http://dx.doi.org/10.1103/PhysRevLett.116.061102}{{\em Phys. Rev.
  Lett.} {\bfseries 116} no.~6, (2016) 061102},
  \href{http://arxiv.org/abs/1602.03837}{{\ttfamily arXiv:1602.03837 [gr-qc]}}.

\bibitem{Aggarwal:2020olq}
N.~Aggarwal {\em et~al.}, ``{Challenges and opportunities of gravitational-wave
  searches at MHz to GHz frequencies},''
  \href{http://dx.doi.org/10.1007/s41114-021-00032-5}{{\em Living Rev. Rel.}
  {\bfseries 24} no.~1, (2021) 4},
  \href{http://arxiv.org/abs/2011.12414}{{\ttfamily arXiv:2011.12414 [gr-qc]}}.

\bibitem{Weber:1967jye}
J.~Weber, ``{Gravitational Radiation},''
  \href{http://dx.doi.org/10.1103/PhysRevLett.18.498}{{\em Phys. Rev. Lett.}
  {\bfseries 18} no.~13, (1967) 498--501}.

\bibitem{Goryachev:2014yra}
M.~Goryachev and M.~E. Tobar, ``{Gravitational Wave Detection with High
  Frequency Phonon Trapping Acoustic Cavities},''
  \href{http://dx.doi.org/10.1103/PhysRevD.90.102005}{{\em Phys. Rev. D}
  {\bfseries 90} no.~10, (2014) 102005},
  \href{http://arxiv.org/abs/1410.2334}{{\ttfamily arXiv:1410.2334 [gr-qc]}}.
  [Erratum: Phys.Rev.D 108, 129901 (2023)].

\bibitem{Aggarwal:2020umq}
N.~Aggarwal, G.~P. Winstone, M.~Teo, M.~Baryakhtar, S.~L. Larson, V.~Kalogera,
  and A.~A. Geraci, ``{Searching for New Physics with a Levitated-Sensor-Based
  Gravitational-Wave Detector},''
  \href{http://dx.doi.org/10.1103/PhysRevLett.128.111101}{{\em Phys. Rev.
  Lett.} {\bfseries 128} no.~11, (2022) 111101},
  \href{http://arxiv.org/abs/2010.13157}{{\ttfamily arXiv:2010.13157 [gr-qc]}}.

\bibitem{Berlin:2023grv}
A.~Berlin, D.~Blas, R.~Tito~D'Agnolo, S.~A.~R. Ellis, R.~Harnik, Y.~Kahn,
  J.~Sch\"utte-Engel, and M.~Wentzel, ``{Electromagnetic cavities as mechanical
  bars for gravitational waves},''
  \href{http://dx.doi.org/10.1103/PhysRevD.108.084058}{{\em Phys. Rev. D}
  {\bfseries 108} no.~8, (2023) 084058},
  \href{http://arxiv.org/abs/2303.01518}{{\ttfamily arXiv:2303.01518
  [hep-ph]}}.

\bibitem{Ito:2022rxn}
A.~Ito and J.~Soda, ``{Exploring high-frequency gravitational waves with
  magnons},'' \href{http://dx.doi.org/10.1140/epjc/s10052-023-11876-2}{{\em
  Eur. Phys. J. C} {\bfseries 83} no.~8, (2023) 766},
  \href{http://arxiv.org/abs/2212.04094}{{\ttfamily arXiv:2212.04094 [gr-qc]}}.

\bibitem{Melissinos:2010zz}
A.~Melissinos and A.~Das, ``{The response of laser interferometers to a
  gravitational wave},'' \href{http://dx.doi.org/10.1119/1.3443566}{{\em Am. J.
  Phys.} {\bfseries 78} (2010) 1160},
  \href{http://arxiv.org/abs/1002.0809}{{\ttfamily arXiv:1002.0809 [gr-qc]}}.

\bibitem{Landau:1975pou}
L.~D. Landau and E.~M. Lifschits, {\em {The Classical Theory of Fields}},
  vol.~Volume 2 of {\em Course of Theoretical Physics}.
\newblock Pergamon Press, Oxford, 1975.

\bibitem{Domcke:2022rgu}
V.~Domcke, C.~Garcia-Cely, and N.~L. Rodd, ``{Novel Search for High-Frequency
  Gravitational Waves with Low-Mass Axion Haloscopes},''
  \href{http://dx.doi.org/10.1103/PhysRevLett.129.041101}{{\em Phys. Rev.
  Lett.} {\bfseries 129} no.~4, (2022) 041101},
  \href{http://arxiv.org/abs/2202.00695}{{\ttfamily arXiv:2202.00695
  [hep-ph]}}.

\bibitem{Ratzinger:2024spd}
W.~Ratzinger, S.~Schenk, and P.~Schwaller, ``{A Coordinate-Independent
  Formalism for Detecting High-Frequency Gravitational Waves},''
  \href{http://arxiv.org/abs/2404.08572}{{\ttfamily arXiv:2404.08572 [gr-qc]}}.

\bibitem{Bringmann:2023gba}
T.~Bringmann, V.~Domcke, E.~Fuchs, and J.~Kopp, ``{High-frequency gravitational
  wave detection via optical frequency modulation},''
  \href{http://dx.doi.org/10.1103/PhysRevD.108.L061303}{{\em Phys. Rev. D}
  {\bfseries 108} no.~6, (2023) L061303},
  \href{http://arxiv.org/abs/2304.10579}{{\ttfamily arXiv:2304.10579
  [hep-ph]}}.

\bibitem{Maggiore:2007ulw}
M.~Maggiore,
  \href{http://dx.doi.org/10.1093/acprof:oso/9780198570745.001.0001}{{\em
  {Gravitational Waves. Vol. 1: Theory and Experiments}}}.
\newblock Oxford University Press, 2007.

\bibitem{Cruise:1983}
A.~M. Cruise, ``{An interaction between gravitational and electromagnetic
  waves},'' \href{http://dx.doi.org/10.1093/mnras/204.2.485}{{\em Monthly
  Notices of the Royal Astronomical Society} {\bfseries 204} no.~2, (1983)
  485}.

\bibitem{Cruise:2000}
A.~M. Cruise, ``{An electromagnetic detector for very-high-frequency
  gravitational waves},'' \href{http://dx.doi.org/10.1093/mnras/204.2.485}{{\em
  Class. Quantum Grav.} {\bfseries 17} (2000) 2525}.

\bibitem{Ejlli:2019bqj}
A.~Ejlli, D.~Ejlli, A.~M. Cruise, G.~Pisano, and H.~Grote, ``{Upper limits on
  the amplitude of ultra-high-frequency gravitational waves from graviton to
  photon conversion},''
  \href{http://dx.doi.org/10.1140/epjc/s10052-019-7542-5}{{\em Eur. Phys. J. C}
  {\bfseries 79} no.~12, (2019) 1032},
  \href{http://arxiv.org/abs/1908.00232}{{\ttfamily arXiv:1908.00232 [gr-qc]}}.

\bibitem{Ringwald:2020ist}
A.~Ringwald, J.~Sch\"utte-Engel, and C.~Tamarit, ``{Gravitational Waves as a
  Big Bang Thermometer},''
  \href{http://dx.doi.org/10.1088/1475-7516/2021/03/054}{{\em JCAP} {\bfseries
  03} (2021) 054}, \href{http://arxiv.org/abs/2011.04731}{{\ttfamily
  arXiv:2011.04731 [hep-ph]}}.

\bibitem{Domcke:2020yzq}
V.~Domcke and C.~Garcia-Cely, ``{Potential of radio telescopes as
  high-frequency gravitational wave detectors},''
  \href{http://dx.doi.org/10.1103/PhysRevLett.126.021104}{{\em Phys. Rev.
  Lett.} {\bfseries 126} no.~2, (2021) 021104},
  \href{http://arxiv.org/abs/2006.01161}{{\ttfamily arXiv:2006.01161
  [astro-ph.CO]}}.

\bibitem{Liu:2023mll}
T.~Liu, J.~Ren, and C.~Zhang, ``{Limits on High-Frequency Gravitational Waves
  in Planetary Magnetospheres},''
  \href{http://dx.doi.org/10.1103/PhysRevLett.132.131402}{{\em Phys. Rev.
  Lett.} {\bfseries 132} no.~13, (2024) 131402},
  \href{http://arxiv.org/abs/2305.01832}{{\ttfamily arXiv:2305.01832
  [hep-ph]}}.

\bibitem{Ito:2023fcr}
A.~Ito, K.~Kohri, and K.~Nakayama, ``{Probing high frequency gravitational
  waves with pulsars},''
  \href{http://dx.doi.org/10.1103/PhysRevD.109.063026}{{\em Phys. Rev. D}
  {\bfseries 109} no.~6, (2024) 063026},
  \href{http://arxiv.org/abs/2305.13984}{{\ttfamily arXiv:2305.13984 [gr-qc]}}.

\bibitem{Dandoy:2024oqg}
V.~Dandoy, T.~Bert\'olez-Mart\'\i{}nez, and F.~Costa, ``{High Frequency
  Gravitational Wave Bounds from Galactic Neutron Stars},''
  \href{http://arxiv.org/abs/2402.14092}{{\ttfamily arXiv:2402.14092 [gr-qc]}}.

\bibitem{Raffelt:1987im}
G.~Raffelt and L.~Stodolsky, ``{Mixing of the Photon with Low Mass
  Particles},'' \href{http://dx.doi.org/10.1103/PhysRevD.37.1237}{{\em Phys.
  Rev. D} {\bfseries 37} (1988) 1237}.

\bibitem{Domcke:2023bat}
V.~Domcke, C.~Garcia-Cely, S.~M. Lee, and N.~L. Rodd, ``{Symmetries and
  selection rules: optimising axion haloscopes for Gravitational Wave
  searches},'' \href{http://dx.doi.org/10.1007/JHEP03(2024)128}{{\em JHEP}
  {\bfseries 03} (2024) 128}, \href{http://arxiv.org/abs/2306.03125}{{\ttfamily
  arXiv:2306.03125 [hep-ph]}}.

\bibitem{Berlin:2021txa}
A.~Berlin, D.~Blas, R.~Tito~D'Agnolo, S.~A.~R. Ellis, R.~Harnik, Y.~Kahn, and
  J.~Sch\"utte-Engel, ``{Detecting high-frequency gravitational waves with
  microwave cavities},''
  \href{http://dx.doi.org/10.1103/PhysRevD.105.116011}{{\em Phys. Rev. D}
  {\bfseries 105} no.~11, (2022) 116011},
  \href{http://arxiv.org/abs/2112.11465}{{\ttfamily arXiv:2112.11465
  [hep-ph]}}.

\end{thebibliography}\endgroup

\end{document}